# Effect of lithium bis(oxalato)borate on the Structural, microstructural and electrochemical properties of blend solid polymer electrolyte

Anil Arya[1], Mohd Sadiq[1,2], A. L. Sharma[1]*

[1]*Department of Physical Sciences, Central University of Punjab, Bathinda, Punjab-151001, INDIA*

[2]*Department of Physics, A R S D College, Delhi 11021, India*

E-mail: alsharma@cup.edu.in

**Abstract**

Blend solid polymer electrolyte based on PEO, PVP complexed with lithium LiBOB were synthesized by solution cast technique and impact of LiBOB on the morphology, structure, and electrochemical properties is examined. The XRD and FESEM analysis reveals the enhancement of amorphous content on salt addition. The FTIR spectroscopy evidences the complex formation. The ionic conductivity for optimized system was two order higher than salt free system, i.e. ~$0.5\times10^{-5}$ S cm$^{-1}$ (@RT) and increases to ~0.65 mS cm$^{-1}$ (@100 °C) for O/Li=16 and temperature dependent conductivity follows Arrhenius nature. The high $t_{ion}$ (~0.99) evidences the ionic nature of complexed electrolyte. DSC analysis evidences the suppression of crystallinity and shift of glass & melting temperature toward lower temperature implies the enhancement of amorphous content. Finally, an interaction mechanism is proposed for exploring the ion transport mechanism. The present electrolyte system can be a potential candidate for application in solid-state batteries

**Key Words**: Blend polymer electrolyte; Ionic conductivity; Activation energy; Glass transition temperature; Relaxation time

1. **Introduction**

The lithium-ion batteries (LIBs) technology is only revolutionary technology that encounters the demand of both stationary & dynamic (S&D) energy storage for humankind all over the world in the electric vehicles (EVs), portable electronics and storing of wind/solar energy. The commercial LIBs system consists of two electrodes with a liquid electrolyte. The electrolyte provides medium for shuttling of lithium ions between electrodes. But the liquid electrolyte is not appropriate, subject to safety problems such as poor mechanical strength, low decomposition voltage, instability, dendrite growth formation, high reactivity, flammability, leakage and bulky size. The goal is to replace the liquid section with solid polymer electrolyte seems to be a most promising alternate option that is getting the marvelous consideration from the scientists. One unique advantage of the polymer is that the cell construction in the desired shape becomes easy. Solid polymer electrolytes (SPEs) are a suitable substitute that may overcome the safety issues, suppress dendrite growth and provides acceptable features such as lightweight, flexibility, ease of fabrication, inert toward electrodes, and may prevent from battery instability [1-5]. But the designed SPE must possess through desirable ionic conductivity, broad stability window, and good thermal/chemical stability. In order to achieve above-mentioned parameters, a number of host polymer have been tried and discussed in the previously published report [6]. The prerequisite property of salts such as low lattice energy, smaller cation, and bulky anion need to be taken care during the complexation. The mechanism of polymer salt complexation has been already reported [7] for better clarity. Among previously reported polymer, the polyethylene oxide (PEO) is most popular aspirant due to broad solubility range of salts, ease of dissolution, and fast migration of cation. But the poor mechanical strength, thermal stability and high crystallinity (71 %) hinders its use in the energy storage devices. The crystalline phase is not suitable for ion migration, so amorphous region is desirable as it facilitates the mutual polymer chain motion which supports the cation migration [8-9]. So, concerning this various approaches are adopted such as polymer blending, branching, grafting, cross-linking etc. Out of aforesaid approaches, polymer blending seems more effective, efficient and easy, as it enables us to achieve the properties of an individual component of the polymer using common solvent. One remarkable advantage is that by varying the composition of the constituents, properties can be tailored as per the requirement [10-13].

Various blend polymer system has been investigated such as PEO-PAN, PEO-PVdF, PVA-PEO, PEO-PEG, PVA-PVP, PVC-PEMA, PEO-PVP and PVC-PEO [14-23]. So, Polyvinylpyrrolidone (PVP) is chosen as blend polymer due to high amorphous content associated with it. Beside this, significant ionic conductivity, high solubility in polar solvents, good thermal stability and ability to provide faster ion migration PVP is getting the attention of the researchers [24-27]. Another most promising approach to enhance the conductivity is the use of salt with a large anion as compared to the traditional salts such as $LiClO_4$, $LiPF_6$, $LiAsF_6$, $LiBF_4$ [28-32]. The feasibility of use of large anion is that immobilization of the anion in polymer backbone may be achieved so that only cation behave as active species which is desirable to reduce the concentration polarization. To eliminate the concentration polarization, salt dissociation plays an important role as conductivity is linked directly to the number of charge carriers. So, in the present study salt lithium bis(oxalate) borate "LiBOB" is chosen due to advantages, low cost, high thermal stability ($T_d$=302 °C), environment-friendliness of its decomposition and hydrolysis products (harmless $B_2O_3$ and $CO_2$, vs. toxic and corrosive HF and/or $POF_3$ from the fluorine-containing lithium salts) [33-35].

In the present paper, we have successfully synthesized the blend solid polymer electrolyte (PEO-PVP+LiBOB) and investigated the structural, electrical and dielectric properties with different LiBOB salt content. The structural, microstructural and electrical properties are measured by the X-ray diffraction, Field emission scanning electron microscope, Fourier transforms infrared spectroscopy and the Impedance spectroscopy study respectively. The ionic transference number is measured by the DC polarization technique and Glass transition temperature, crystallinity by the Differential scanning calorimetry. The ion dynamics and dielectric relaxation of prepared polymer electrolytes with frequency variation are explored further by the dielectric analysis. The dielectric analysis enclosed the complex permittivity, loss tangent, complex conductivity, and modulus formalism and all the experimental plots were simulated with the theoretical fitting in the complete frequency window. Finally, a correlation between the various relaxation time of the blend polymer electrolyte system has been established.

2. **Experimental**
*2.1. Materials*

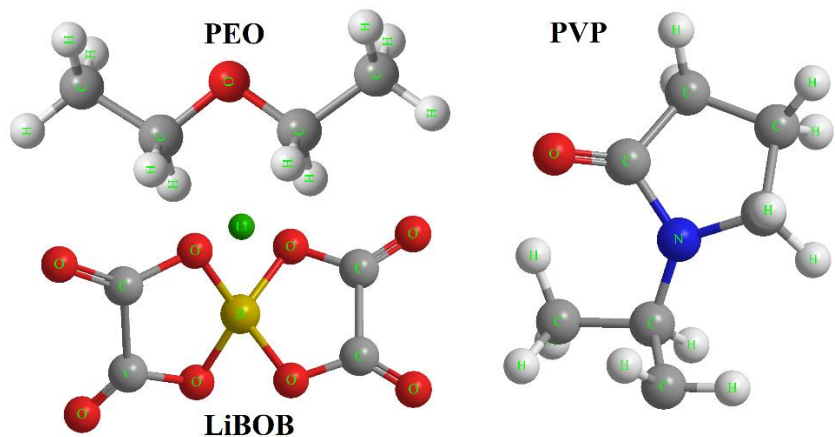

Figure 1. Structure unit of PEO, PVP and salt LiBOB.

The polymer, PEO (av. Mol. Wt.; 200,000, Sigma), PVP (av. Mol. Wt.; 40000, Sigma), LiBOB (Mol. Wt. 167.95 g/mol, Sigma) and methanol (Loba) were used as received. The chemical structure of all material is displayed in pictorial form as shown in Fig. 1.

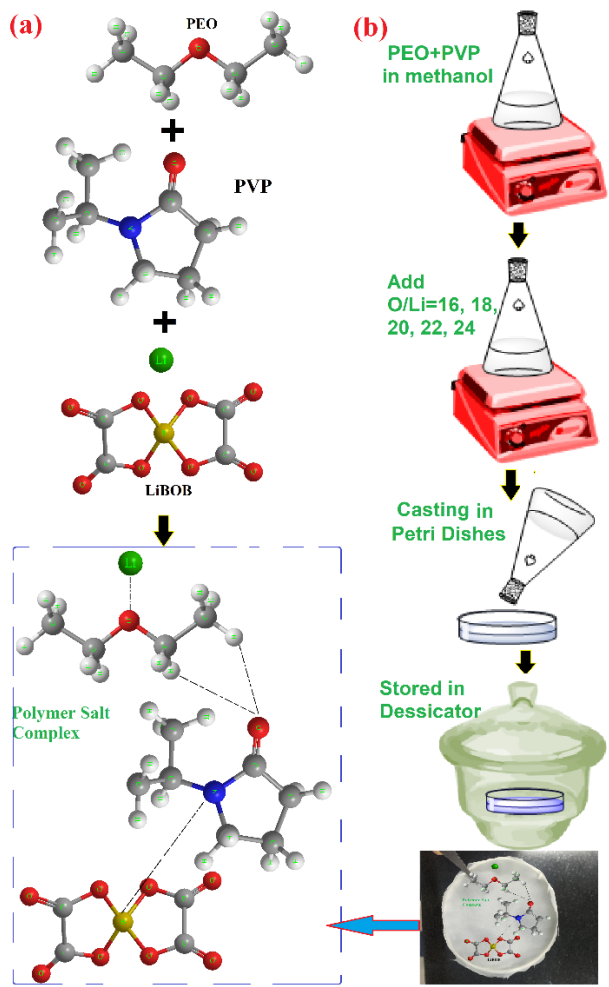

**Figure 2.** **(a)** Representation of blend formation and cation coordination, **(b)** flow chart of solution cast technique.

An appropriate amount of the polymer (80:20) and salt (O/Li$^+$ = 16, 18, 20, 22, 24) were mixed methodically using the common solvent methanol by standard solution cast technique. Appropriate amount of PEO and PVP were added in methanol (20 ml) and kept for 10 minutes for swelling of polymer chains followed by 4 hours stirring. Then the appropriate stoichiometric ratio of salt (O/Li$^+$ = 16, 18, 20, 22, 24) were added to the homogeneous solution and stirred for 12 hours before casting in the Petri-dishes. Then after drying at room temperature first and then in a vacuum oven (for 24 hours), the prepared film is peeled off from Petri-dishes and stored in a desiccator to avoid contamination and for further relevant characterizations. Figure 2 a & b displays the interaction mechanism of polymer salt complex and diagram of solution cast technique.

### *2.2. Characterizations and Basic Background*

#### *2.2.1. X-ray diffraction*

The interlayer spacing (*d*) and the interchain separation (R) was investigated by the (XRD) (Bruker D8Advance) performed and recorded with Cu-K$_\alpha$ radiation ($\lambda$=1.54 Å) in the Bragg's angle range (*2θ*) from 10° to 60° [36].

Interlayer spacing (*d*)   $2d\sin\theta = n\lambda$   (1)

and

Interchain separation (*R*)   $R = 5\lambda/8\sin\theta$   (2)

*Field emission scanning electron microscopy (FESEM)*

The surface morphology and the topography were investigated by the FESEM (Carl Zeiss). The samples were taken in a high vacuum after sputtering with gold in order to prepare conductive surfaces.

#### *2.2.2. Fourier Transform Infrared Spectroscopy*

The presence of the various interaction in the blend solid polymer electrolyte was evidenced by the FTIR (Bruker Tensor 27, Model: NEXUS–870) in absorbance mode over the wavenumber region from 600 to 3500 cm$^{-1}$ (resolution of 4 cm$^{-1}$).

#### *2.2.3. Impedance Study*

The ionic conductivity was measured by impedance spectroscopy (IS) in the frequency range of 1 Hz to 1 MHz using the CHI (Model:760 ; USA) electrochemical analyzer. An AC sinusoidal signal of 20 mV was applied to the cell configuration SS|SPE|SS where solid polymer electrolytes films were sandwiched between two stainless steel electrodes. The intercept between the semi-circle at high frequency and tilted spike at low frequency were taken as the bulk resistance ($R_b$). The bulk conductivity ($\sigma$) value was obtained using equation 1:

$$\sigma_{dc} = \frac{1}{R_b}\frac{t}{A} \quad (3)$$

Where '*t*' is thickness (cm) of polymer film, R$_b$ is bulk resistance ($\Omega$) and A is area (cm$^2$) of working electrode. The temperature dependent conductivity was studied in the temperature range from 40 °C to 100 °C with a temperature difference of 10 °C (Temperature Controller; Marine India). All the samples were kept for 20 minutes to attain the thermal equilibrium before each measurement. The thermal activation energy for ionic transport was estimated from the slope of the linear fit of the Arrhenius plot. The linear variation in log ($\sigma$/Scm$^{-1}$) vs. 1000/*T* plot suggests a thermally activated process represented by Arrhenius equation 4;

$$\sigma = \sigma_o \, exp(-E_a/kT) \qquad (4)$$

Where, $\sigma_o$ is the constant pre-exponential factor and $E_a$ is the activation energy. The parameter T stands for the absolute temperature and $k$ for the Boltzmann constant.

*2.2.4. Differential Scanning calorimetry Study*

To evaluate the glass transition temperature, melting temperature and the crystallinity DSC (DSC-Sirius 3500) was performed with a heating rate of 10 °C min$^{-1}$ under an N$_2$/Ar atmosphere. Solid polymer electrolyte films with the weight of 8-10 mg were sealed in aluminum pans, and an empty sealed aluminum pan was used as a reference. The crystallinity was calculated using the following equation 5;

$$X_c = \frac{\Delta H_m}{\Delta H_m^o} \times 100 \qquad (5)$$

Where $\Delta H_m$ is the melting enthalpy obtained from the DSC measurement and $\Delta H_m^o$ is the melting enthalpy of pure 100% crystalline PEO (188 J/g) [8, 37].

*2.2.5. Ion transference number study*

The total ionic transference number ($t_{ion}$) was obtained by placing polymer electrolyte film between stainless steel (SS) blocking electrodes and a fixed dc volatge of 10 mV was applied across the *SS/SPE/SS* cell (@ 40 ºC). Ion transference numbers of the solid polymer electrolytes was evaluated using equation 6 on *SS/SPE/SS* cell:

$$t_{ion} = \left(\frac{I_t - I_e}{I_t}\right) \times 100 \qquad (6)$$

The samples are labeled as PP, PP16, PP18, PP20, PP22, and PP24 for blend solid polymer electrolyte films with $Ö/Li^+ = 0, 16, 18, 20, 22, 24$ of LiBOB respectively for further investigation.

## 3. Results and Discussion
### 3.1. X-ray diffraction (XRD) study

The XRD spectra of the PEO-PVP blend polymer electrolyte with different salt content are presented in Figure 3. The most intense peaks located at 19.36º and 23.72º are associated with the PEO, corresponding to the plane (120) & (112) and represent the crystalline phase of the PEO due to the presence of strong hydrogen bonding. Then a low-intensity peak at 15º, 26º and 27º of PEO are associated with the plane (013), (222), (111) respectively [38]. The presence of a small peak in the blend polymer located at 13º is attributed to the amorphous nature of PVP. Further addition of salt leads to the disappearance of peak and it confirms the multiphase (both crystalline and amorphous phase). The salt addition in the blend polymer electrolyte alters the peak intensity and position that confirms the formation of polymer salt complex due to polymer---salt interaction. The decrease of intensity associated with the most intense peak of PEO suggests the decrease of the crystallinity. It may be due to the interaction of the cation with the polymer chain that disrupts the ordered arrangement of the polymer chain and strengthens the amorphous content. Further, the disappearance of the salt peak confirms the complete dissolution of the salt in the polymer matrix and good miscibility [39]. The d-spacing between the diffraction planes was obtained using the equation 1, and interchain separation (*R*) using the equation 2 and determined values are shown in Table 1.

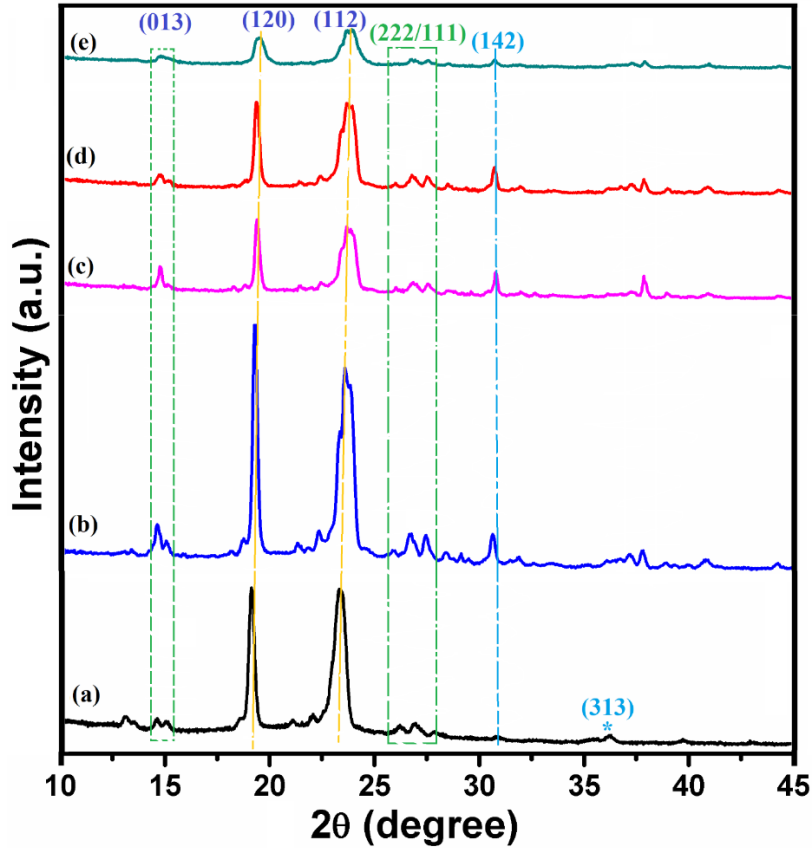

**Figure 3.** XRD patterns of blend polymer electrolyte films with LiBOB (O/Li)=(a) 0, (b) 24, (c) 22, (d) 20, (e)16.

It may be observed from the Table 1 that the *d*-spacing and the interchain separation are varying slightly on the addition of the salt that indicates that the covalent bonding between the polymer chain is disrupted that results in the lowering of the polymer chain viscosity and hence the faster ion migration is achieved [7].

**Table 1.** Values of Bragg's angle 2θ, Basal spacing (d), and inter-chain separation (R) corresponding to 120 and 112 reflection planes of PEO in blend solid polymer electrolyte films.

| Sample Code | PEO 120 reflection plane parameters | | | PEO 112 reflection plane parameters | | |
|---|---|---|---|---|---|---|
| | 2θ (degree) | d (Å) | R (Å) | 2θ (degree) | d (Å) | R (Å) |
| PP | 19.05 | 4.65 | 5.81 | 23.06 | 3.85 | 4.81 |
| PP24 | 18.95 | 4.67 | 5.84 | 23.08 | 3.84 | 4.80 |
| PP22 | 18.96 | 4.67 | 5.84 | 23.14 | 3.83 | 4.79 |
| PP20 | 18.93 | 4.68 | 5.85 | 23.15 | 3.83 | 4.79 |
| PP16 | 18.99 | 4.67 | 5.83 | 23.19 | 3.83 | 4.78 |

### 3.2. Field Emission Scanning Electron Microscopy (FESEM) study

The Field emission scanning electron microscopy (FESEM) technique is performed to investigate the morphology and the elemental composition by the Energy-dispersive X-ray spectroscopy (EDS). Figure 4 displays the FESEM micrographs of the blend polymer electrolyte films. Figure 4 a showed the micrograph of the pure PEO and rough/rigid structure indicates the crystalline nature of the PEO. When PVP is added in the PEO then the surface roughness is improved and the modified surface is observed that indicates the blend formation (Figure 4 b).

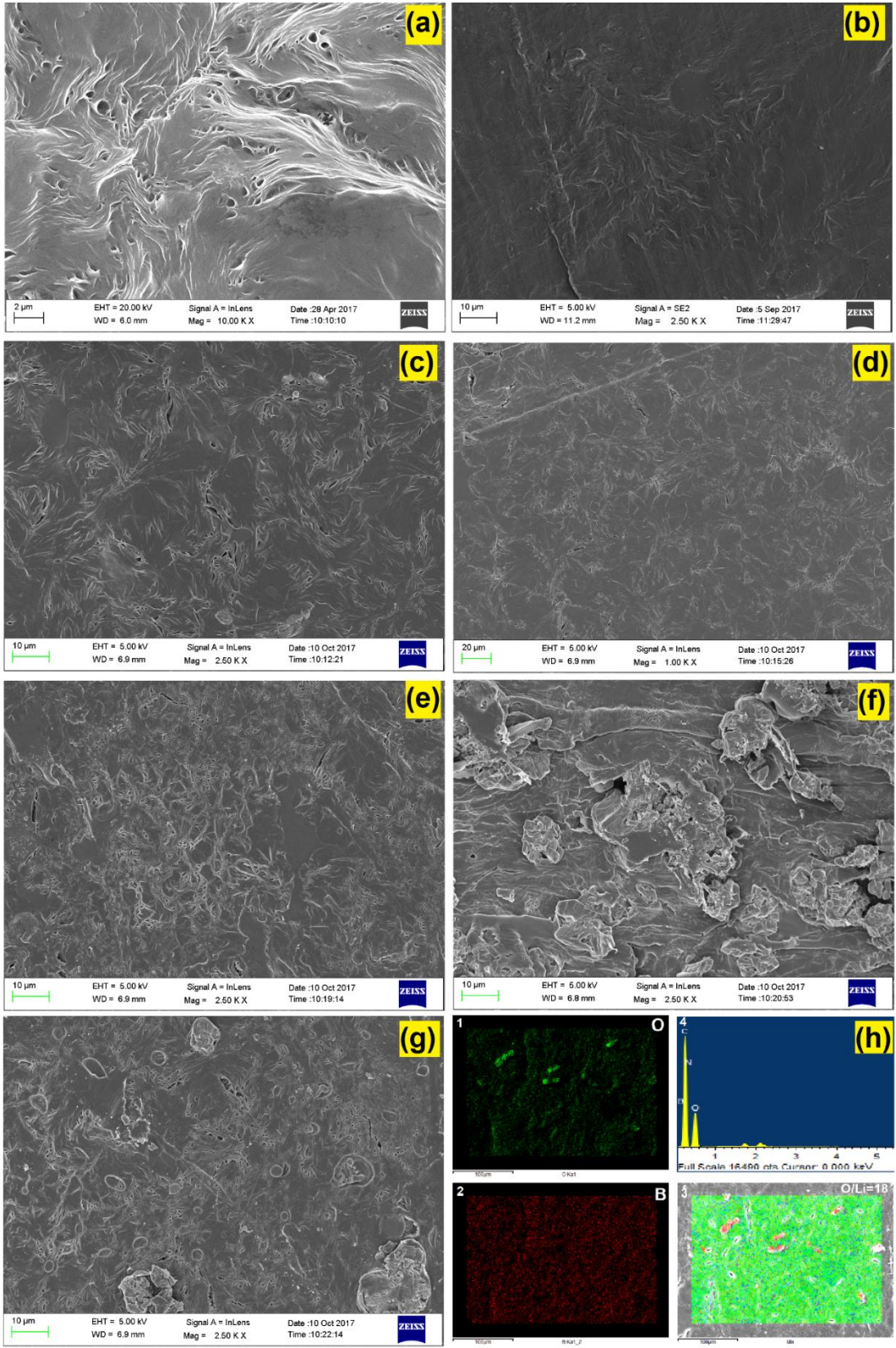

**Figure 4.** FESEM micrographs of the (a) PEO, (b) blend polymer electrolyte (PEO-PVP), (c-g) Blend PEO-PVP with O/Li= 0, 16, 18, 20, 22, 24 of LiBOB and (h) elemental mapping and EDX spectra of the O/Li=16.

The absence of the cracks and formation of the smooth surface indicates the polymer miscibility due to hydrogen bonding between the polymers. Further addition of the salt modifies the surface morphology and indicates the formation of the polymer salt complex via the coordinating interaction of the lithium ion with ether group of PEO [40]. The uniform distribution of the salt in the polymer matrix enhances the amorphous phase and is evidenced by the XRD analysis. For better visibility of the salt dissociation, EDS has been performed and it confirms the uniform presence of the salt in the polymer matrix. Further, the elemental composition of the O and B of the salt suggest complete dissolution of the salt (Figure 4 h). It might be concluded that the addition of the salt alters the ordered arrangement of the polymer chains and smoother surface facilitate the faster ion migration.

### 3.3. Fourier Transform Infrared Spectroscopy (FTIR) study

The presence of various functional groups in the polymer salt complex comprising of PEO-PVP+LiBOB is observed by the FTIR spectra. Figure 5 displays the FTIR spectra of the blend polymer (a) and complexed with different salt content (b-f). The spectral region in the wavenumber range 600 $cm^{-1}$ to 1600 $cm^{-1}$ consists of the $CH_2$ rocking, C-O st., $CH_2$ asymm. rocking, C-O-C st. symm/asymm., $CH_2$ twisting and wagging mode. The band observed at the 846 $cm^{-1}$ is associated with the $CH_2$ rocking mode and some contribution from the C-O stretching mode. The band located at the 958 $cm^{-1}$ is attributed to the symm./asymm. C-O-C stretching mode. The band observed at the 1280 $cm^{-1}$, 1340 $cm^{-1}$ and 1480 $cm^{-1}$ corresponds to the asymm. $CH_2$ twisting, $CH_2$ bending/wagging, and C-H bending respectively. Addition of the salt in the blend polymer electrolytes alters the peak position and intensity that indicates that the salt plays an effective role in altering the polymer arrangement [40, 41]. The C-O-C st. observed at 1100 $cm^{-1}$ indicates the amorphous nature and after addition of salt peak get broadened and is asymmetrical that strongly evidences that the cation is going to coordinate with the ether group of PEO than the PVP. This also confirms that the PEO has a high probability of attracting the cation and ether group acts as a strong electron donor. The overall effect is the disruption of the ordered arrangement of the polymer chains and amorphous content is increased and is in agreement with the XRD results which show a reduction of the peak intensity and hence the crystallinity. It suggests that the addition of the salt modifies the polymer chain arrangement via the coordinating interaction between the cation and the electron rich group of the PEO. The ether group promotes the salt dissociation in the blend polymer matrix. This interaction alters the peak position and intensity which indicates that the polymer crystallinity is disrupted.

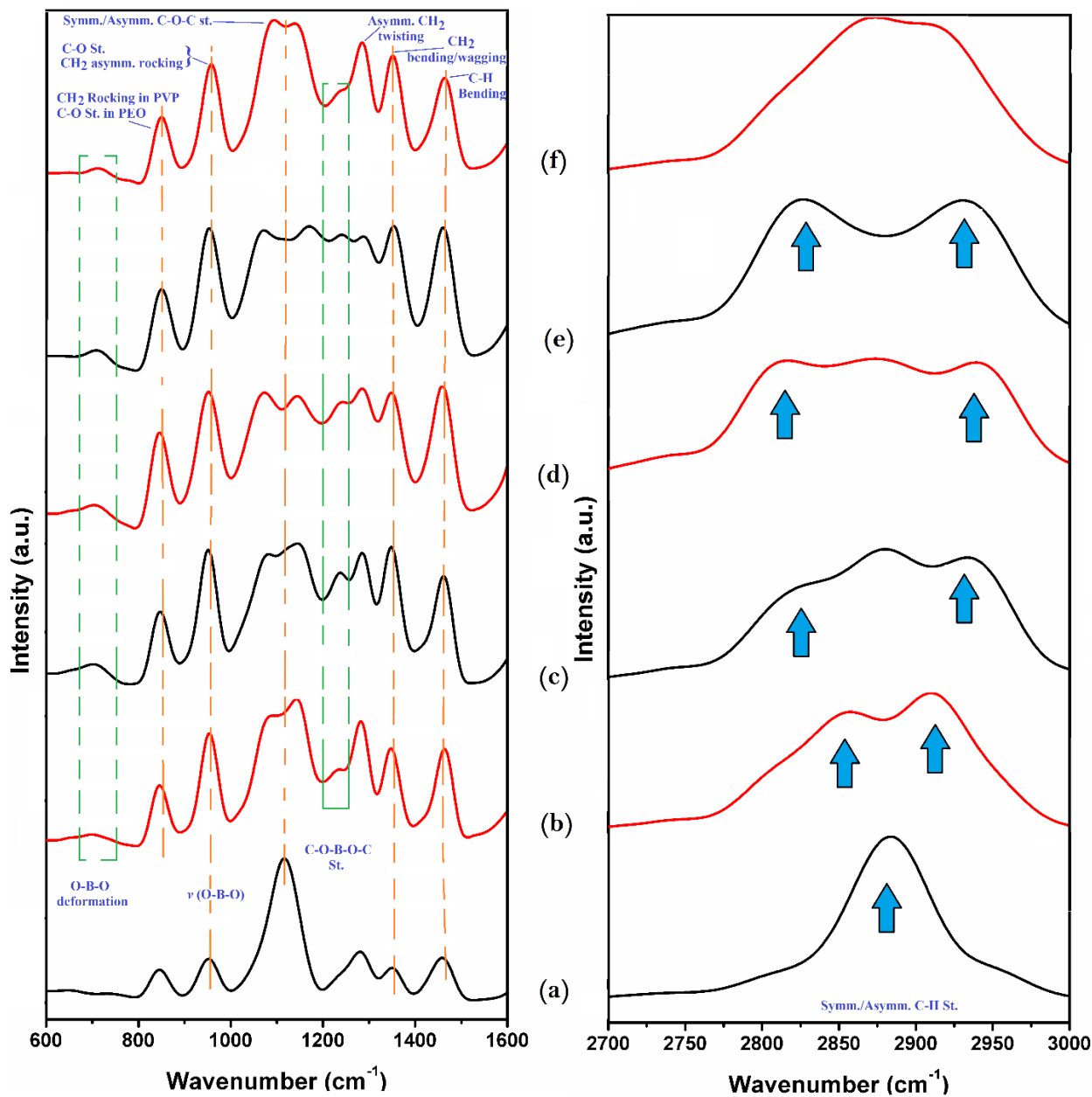

**Figure 5.** FTIR spectra of blend polymer electrolyte films with O/Li ratio (a) 0, (b) 16, (c) 18, (d) 20, (e) 22, (f) 24

Now, the presence of BOB⁻ group in LiBOB has led to the 10 Raman and 7 IR active vibrational modes. The main band is observed (Figure 5) at 775, 985, 1089 and 1400 cm$^{-1}$ attributed to the O-B-O deformation, ν(O-B-O)+ δ (O-C-O), O-B-O symmetric stretch and B-O extra-ring stretch, respectively [42-44]. The change in the anion peak with an increase of salt content is observed in terms of the change in the peak shape and the negligible change in position in the blend polymer electrolyte. Figure 6 depicts the deconvolution spectra of the anion peak in the wavenumber region of 1700-1850 cm$^{-1}$ and peaks are associated with the C=O stretch of the LiBOB. Another important region is the 2800-2900 cm$^{-1}$ and is associated with the symm/asymm. C-H stretching of the PEO. For the blend polymer, a symmetrical peak is observed and the addition of salt disrupts the symmetry of the peak and two peaks can be observed that suggests that the salt plays an effective role in altering the blend polymer electrolyte (Figure 6 b-f). So, it can be summarized that the incorporation of the salt in the blend polymer electrolyte alters the structure of polymer matrix and interaction between the polymer-salt are modified. It can be concluded from the above that the polymer salt complexation occurs which is evidenced by the XRD and FESEM analysis. Further investigation regarding the modification of the structure is analyzed by evaluating the glass transition temperature, melting temperature and crystallinity in the upcoming section.

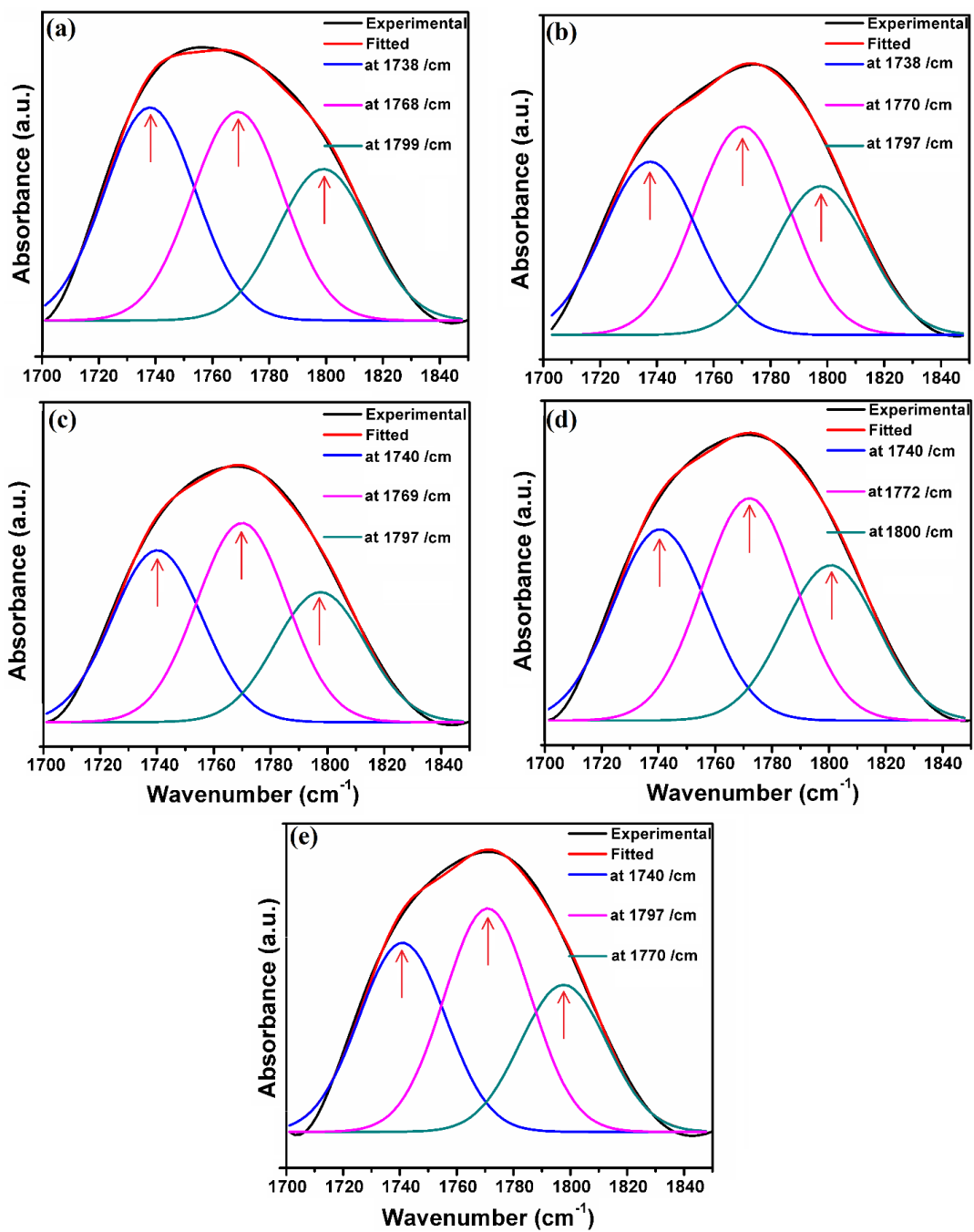

**Figure 6.** FTIR deconvolution of the C=O peak of salt for O/Li ratio (a) 16, (b) 18, (c) 20, (d) 22, (e) 24.

### 3.4. Impedance Spectroscopy (IS)

The impedance spectroscopy has been performed by sandwiching the polymer electrolyte between two stainless steel blocking electrodes. Then an ac signal (~50 mV) is applied across the SS electrodes and the plot of imaginary part vs. real part of impedance is observed in the frequency range 1 Hz to 1 MHz. Figure 7 a summarizes the impedance plot which comprises of the two regions, one semicircle at the high frequency which is due to the bulk nature of the sample (associated with ion migration) followed by the tilted spike of the low frequency that is associated with the electrode double layer capacitance of the electrodes at interface [40, 45-47]. The minima in the imaginary part of the impedance are used to obtain the bulk resistance ($R_b$). In addition, it is also seen from the plot that bulk resistance decreases with the addition of the salt in the polymer matrix. The lowest value was recorded for the PP16. The addition of salt provides more free charge carriers in the polymer matrix and available charge carriers participate in the conduction hence increase in the conductivity is observed (Figure 7 a). 2.

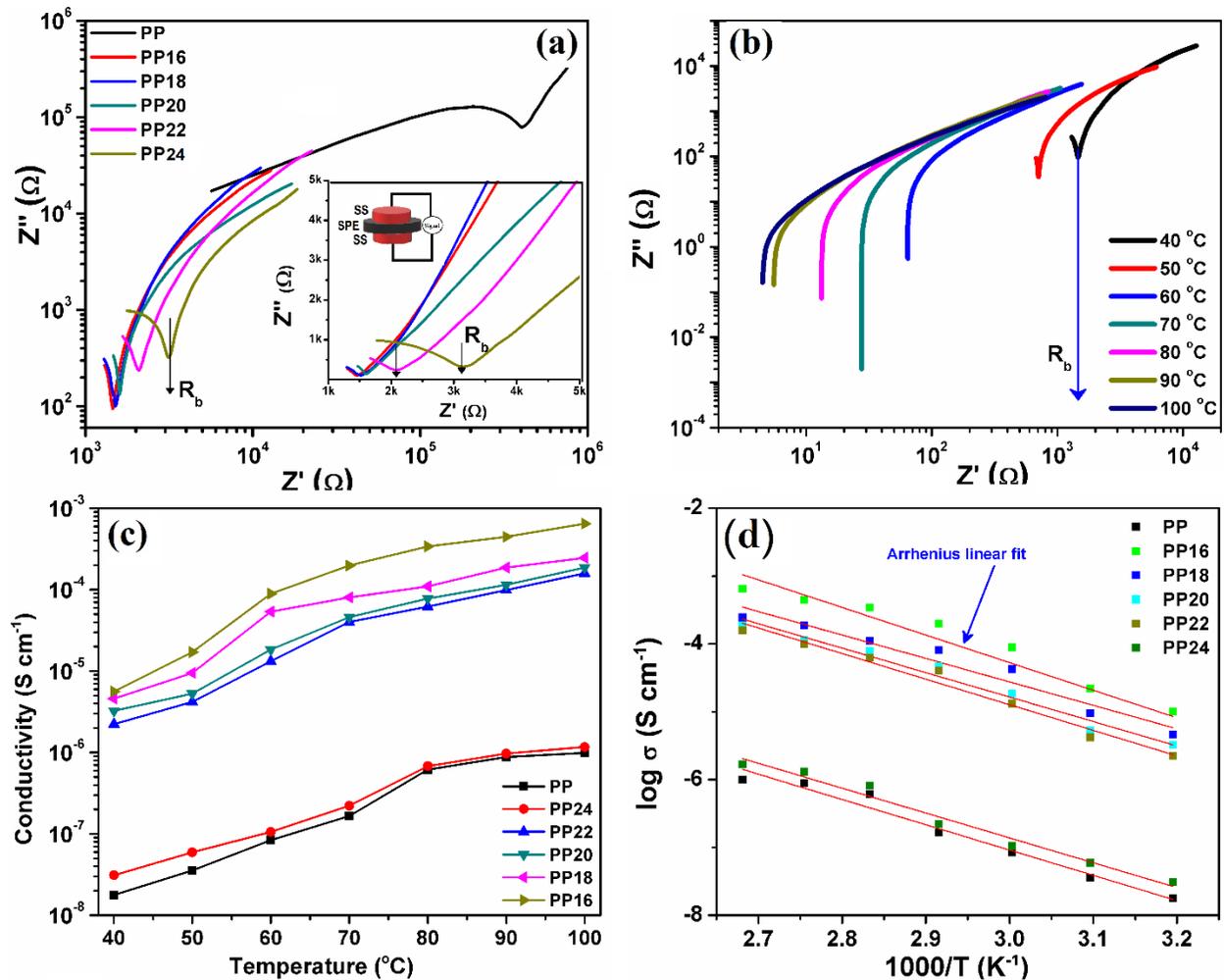

**Figure 7.** (a) Impedance plot of blend polymer electrolyte with different salt content, (b) temperature variation of PP16 blend polymer electrolyte, (c) Variation of conductivity against temperature for different salt content, (d) Reciprocal temperature dependence of the conductivity for blend polymer electrolyte.

So, in comparison to the polymer electrolyte without salt, blend polymer electrolyte with salt displays increased conductivity which is possible only when the polymer matrix have amorphous content, faster ion mobility and enhanced polymer flexibility. The conductivity is calculated using the equation 3 and summarized in Table 2. The highest ionic conductivity of $0.48 \times 10^{-5}$ S cm$^{-1}$ at 40 °C is observed for PP16. The ionic conductivity value was higher than the PEO-PVP+LiClO4 based system which displays a conductivity of $0.37 \times 10^{-5}$ S cm$^{-1}$ at 40 °C [48].

Now the temperature dependent conductivity is also evaluated for PP16 and it is observed that the bulk resistance decreases with the increase of temperature. It indicates that the activation of charge carriers occurs which enhance the rate of ion migration from one coordinating site to another [49]. It suggests that the increase of temperature enhance the polymer flexibility and ion mobility which is supported by the faster segmental motion of polymer chain (Figure 7 b). Figure 7 c depicts the conductivity variation for all blend polymer system with the increase of the temperature. As the temperature increases then the increase in the conductivity is observed and is due to following reasons, (i) increased polymer flexibility, (ii) increase in free volume, (iii) faster segmental motion of polymer chain, and (iv) reduced polymer chain viscosity.

**Table 2.** Different contributions of electrical conductivity (ionic & electronic) and transference number for blend solid polymer electrolyte.

| Sample Code | Transference Number @ 40 °C | Electrical Conductivity (S cm$^{-1}$) @ 40 °C | Electrical Conductivity (S cm$^{-1}$) @ 100 °C | Electronic Conductivity (S cm$^{-1}$) | Ionic Conductivity (S cm$^{-1}$) |
|---|---|---|---|---|---|
| PP | - | $0.17 \times 10^{-7}$ | $0.98 \times 10^{-6}$ | - | - |
| PP24 | 0.95 | $0.31 \times 10^{-7}$ | $0.11 \times 10^{-5}$ | $0.11 \times 10^{-8}$ | $0.29 \times 10^{-7}$ |
| PP22 | 0.95 | $0.22 \times 10^{-5}$ | $0.15 \times 10^{-3}$ | $0.12 \times 10^{-8}$ | $0.21 \times 10^{-5}$ |
| PP20 | 0.98 | $0.35 \times 10^{-5}$ | $0.18 \times 10^{-3}$ | $0.70 \times 10^{-7}$ | $0.34 \times 10^{-5}$ |
| PP18 | 0.99 | $0.45 \times 10^{-5}$ | $0.24 \times 10^{-3}$ | $0.45 \times 10^{-7}$ | $0.45 \times 10^{-5}$ |
| PP16 | 0.99 | $0.50 \times 10^{-5}$ | $0.65 \times 10^{-3}$ | $0.48 \times 10^{-7}$ | $0.48 \times 10^{-5}$ |

Figure 7 d depicts the temperature dependence of the ionic conductivity and follows the Arrhenius behavior (Eq. 4). As the temperature is increased then the ion migration from one coordination site to another occurs fast due to thermal activation and is associated with the lowering of the activation energy. The solid line in the plot shows the Arrhenius fit. The lowering of the activation energy evidences the significant increase in the ionic conductivity as compared to the blend polymer electrolyte without salt [40]. Lowering in the activation energy suggests the availability of a favorable path for cation migration and conductivity reaches to $0.65 \times 10^{-3}$ S cm$^{-1}$ at 100 °C (for PP16) which lies within the functioning range of the solid-state battery. So, it may be suggested that both salt concentration and temperature are the key players that control the ion transport in the blend polymer matrix.

Above investigations support the FTIR results which evidence the lowering of the interaction between cation/anion and salt dissociation increases. This separates the cation and anion. So, the amorphous phase in the blend polymer matrix allows the cation migration supported by the segmental motion of polymer chains while anion is attached to the polymer backbone. This leads to increase in the conductivity and polymer flexibility while displays reduction of crystallinity as evaluated from XRD. So, the large anion of the LiBOB facilitate the improved electrical properties as expected. To further support the ionic conductivity, which is evidenced by the shift of $T_g$ toward lower temperature as studied from DSC in forthcoming section.

**3.5. Differential Scanning Calorimetry (DSC)**

Figure 8 illustrates the DSC thermograms of the blend polymer electrolyte films with different salt content. DSC was used to calculate the glass transition temperature ($T_g$), crystallinity ($X_c$) and the melting temperature ($T_m$). From the Figure 8, the onset $T_g$ was observed in the range of 45-60 °C. As it is well known that the $T_g$ of the PEO is -57.6 °C and of PVP is 69 °C. While the present system displayed only single $T_g$ between the individual polymer that confirms the miscibility of the polymer. The motion of polymer chains is the key player that facilitates the cation migration and is evidenced by the glass transition temperature value. Addition of salt lowers the $T_g$ and it indicates that the polymer flexibility increases due to the interpenetration of the anion between the polymer chains (Table 3). The increased flexibility is an indication of the amorphous phase that is desirable for the fast solid state ionic conductor [45, 50]. The relative crystallinity was calculated from the DSC thermogram using the Eq .(5). It is observed that the addition of salt lowers the crystallinity. The lowering of the crystallinity is attributed to the coordination interaction between the polymer chain and the cation ($Li^+$). As FTIR also evidenced the cation coordination with the electron rich group of the polymer chain and anion with polymer backbone. So, both cation and anion together alter the polymer chain arrangement and more free volume is available for the cation migration [40, 50].

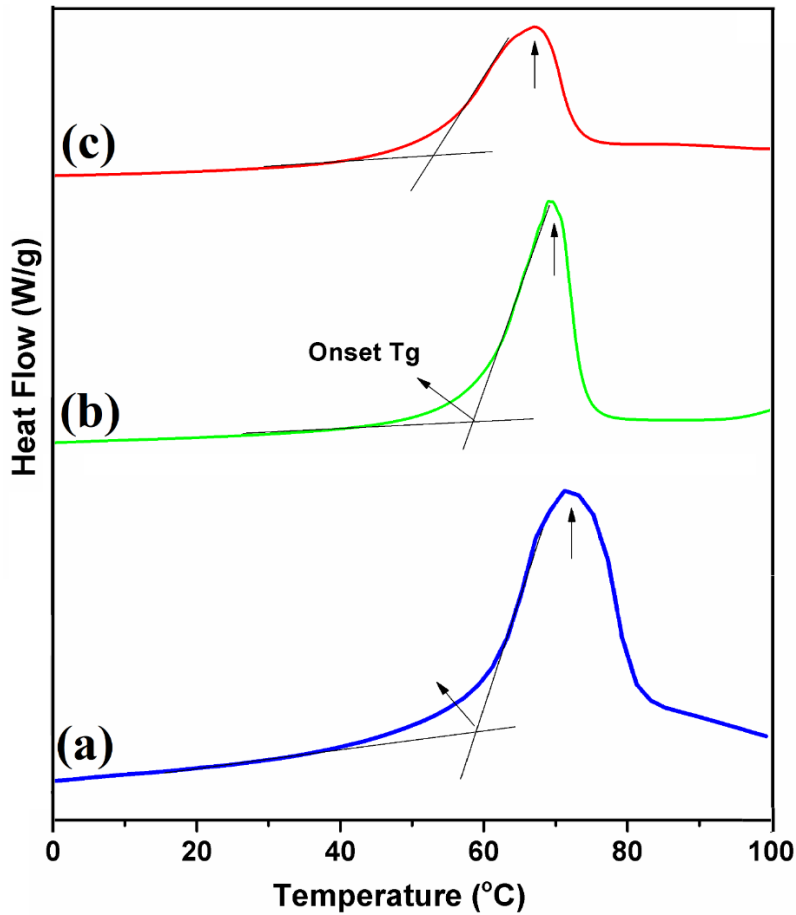

**Figure 8.** DSC thermograms of (a) PEO-PVP, (b) O/Li=20 and (c) O/Li=16.

Table 3. Glass transition temperature, crystallinity and the melting temperature of the blend polymer electrolyte and blend polymer electrolyte with salt content.

| Sample Code | $T_g$ (onset, °C) | $T_m$ (°C) | $X_c$ (%) |
|---|---|---|---|
| PP | 59.78 | 71.7 | 66.75 |
| PP20 | 57.89 | 69.2 | 53.29 |
| PP16 | 53.17 | 67.1 | 49.09 |

The decrease of crystallinity is linked directly with the polymer flexibility and the segmental motion of the polymer chains. The reduction of crystallinity is also observed from the XRD analysis. The melting temperature peak also shows shifts toward the lower temperature on the addition of the salt. It may be attributed to the lowering of the crystallinity and chain stiffness. The blend polymer electrolyte (BPE) complexed with salt displays the lowering of glass transition temperature, melting temperature, and crystallinity. This decrease is attributed to the cation interaction with the ether group of PEO which disrupts the ordered arrangement of the polymer chains. Observed decrease in all three parameters suggest that the investigated system has faster ion migration rate which is seen in terms of the increase in conductivity and transference number.

### 3.6. Transference Number

As a polymer electrolyte system both ions/elecrons contribute to the total conductivity. So, to separate out the ionic and electronic contribution, a well established DC polarization method has been performed by placing the solid polymer electrolyte in between the stainless steel (SS) electrodes and a signal of 20 mV used to be applied.

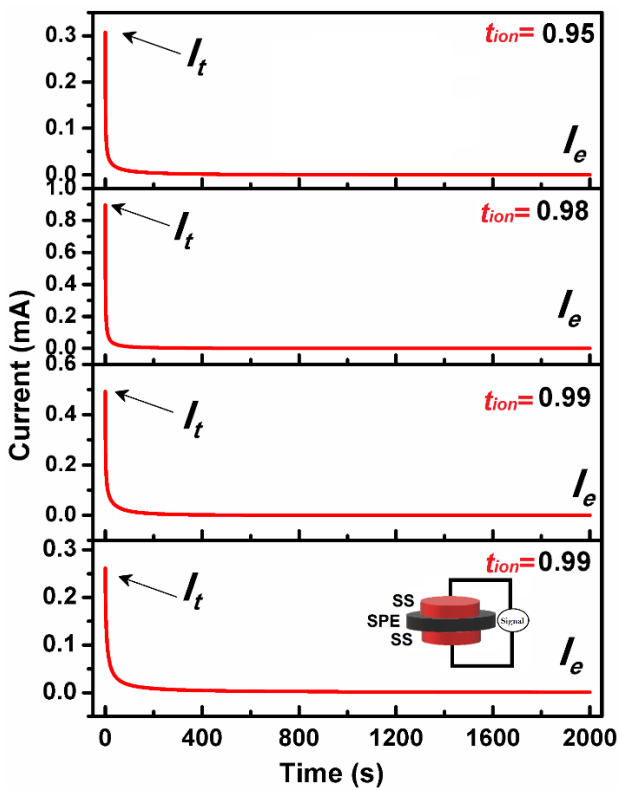

**Figure 9.** Ion transference number of blend polymer electrolyte films (a) O/Li=16, (b) O/Li=18, (c) O/Li=20 and O/Li=22.

Figure 9 displays the plot of the polarization current vs. time for all systems. All the plot depicts the same trend of decrease of the current with time followed by a steady state for a long time. Actually, on the application of the filed along with the cell configuration, all ions are accumulated on the interface due to polarization and further ion migration is blocked [51]. So, only active species now left is with electrons only. All the systems show a high value of ion transference number (~99 %) obtained by Eq. (6) in Table 2. This implies that the total conductivity in the present system is mainly due to flow of ions with negligble electron contribution.

### 4. Self-Proposed Mechanism for Ion Transport

Figure 10 displays the ion transport mechanism of the investigated system. Figure 10 a, depicts the PEO chain and the PVP chain. The blend formation is depicted in Fig. 10 a, where the oxygen of the PVP chain interacts with the methyl group of the PEO chain via hydrogen bonding. The blend formation at some level alters the polymer chain arrangement which facilitates the more free space for ion migration. Now, when salt is added to the blend polymer matrix, then the salt gets dissociated in cation & anion. From here, two possibilities arise for the interaction of cation, (i) interaction with ether group of PEO and (ii) with the oxygen PVP. But, from the FTIR spectra, it was evidenced that the cation coordination seems to be with the ether group of the PEO.

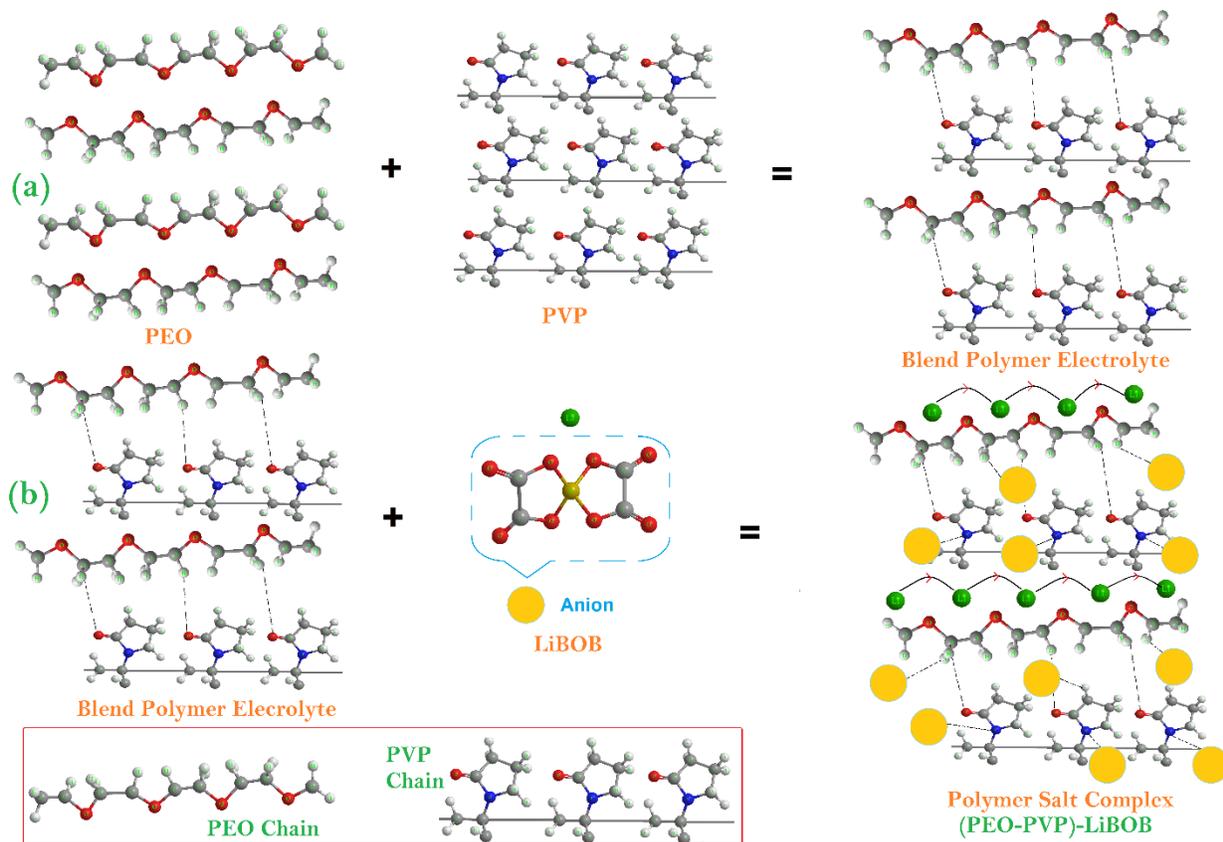

**Figure 10.** Model for the Lithium ($Li^+$) migration, (a) blend formation interaction mechanism and (b) Polymer salt complex formation.

As it is evidenced by FTIR results, the principal peak of the PEO at 1100 cm$^{-1}$ shows asymmetry on the addition of salt which is clear indication that the ether group of PEO is the preferable site of coordination. Now, another species remains is an anion. As anion is of large size so it gets attached to the polymer backbone and remains in the immobilized state (Figure 10 b). The overall effect of the salt addition is disruption of the polymer chain arrangement that indicates the enhancement of the amorphous content as evidenced by the XRD and DSC results. It can be summarized that the cation migration occurs via the coordinating sites provided by PEO while PVP supports the backbone of the PEO. Now, when the salt content is varied, a number of charge carriers participate in conduction that results in improvement in electrical conductivity. Further, the decrease of glass transition temperature and crystallinity with the addition of salt suggests the increase of polymer flexibility/faster segmental motion of polymer chain which leads to faster ion migration.

5. **Conclusions**

In summary, solid blend polymer electrolyte based on PEO-PVP and LiBOB were synthesized by the standard solution cast technique. The structural and morphological analysis was performed by the X-ray diffraction and Field emission scanning electron microscope. The reduction of peak intensity on the addition of the salt evidences the lowering of crystallinity on the addition of salt. The surface morphology was altered on the addition of the salt that indicates the enhancement of the amorphous content. The ionic conductivity was obtained by the Impedance spectroscopy technique and was of the order of the $10^{-5}$ S/cm. It suggests that the large anion size effectively enhance the ionic conductivity. The temperature dependence of the ionic conductivity follows Arrhenius behavior and activation energy decreases with the addition of the salt. The increased conductivity with temperature was the result of increased polymer flexibility and lowering of activation energy. The high value of the ionic transference number confirms the ionic nature of the prepared polymer electrolyte system. The lowering of the glass transition temperature and the melting temperature indicates the enhancement of the amorphous content and support the enhancement of ionic conductivity value. On the basis of information obtained from different characterization techniques, a self-convincing mechanism has been proposed for the better visibility to the readers. All the above-stated properties make the present polymer electrolyte a potential candidate as an electrolyte cum separator of lithium-ion battery.

**Acknowledgment**

One of the author AA is thankful to the Central University of Punjab for providing fellowship and partial funding from the UGC Startup Grant (GP-41).